\documentclass[reprint,superscriptaddress,floatfix,amsmath,amssymb,aps,prl]{revtex4-1}

\usepackage{graphicx}
\usepackage{dcolumn}
\usepackage{bm}
\usepackage{textcomp}

\usepackage{xcolor}
\definecolor{prlblue}{rgb}{0.176, 0.152, 0.57}
\usepackage[colorlinks=true, linkcolor=black, citecolor=prlblue, filecolor=black, urlcolor=prlblue]{hyperref}

\definecolor{light-gray}{gray}{0.8}
\usepackage[mathlines]{lineno}

\begin{document}

\title{Longitudinally Resolved Measurement of Energy-Transfer Efficiency \\ in a Plasma-Wakefield Accelerator}

\author{L. Boulton}
\email{lewis.boulton@desy.de}
\affiliation{Deutsches Elektronen-Synchrotron DESY, Notkestr.~85, 22607 Hamburg, Germany}
\affiliation{SUPA, Department of Physics, University of Strathclyde, Glasgow, United Kingdom}
\affiliation{The Cockcroft Institute, Daresbury, United Kingdom}
\author{C.~A.~Lindstr{\o}m}
\affiliation{Deutsches Elektronen-Synchrotron DESY, Notkestr.~85, 22607 Hamburg, Germany}
\author{J.~Beinortaite}
\affiliation{Deutsches Elektronen-Synchrotron DESY, Notkestr.~85, 22607 Hamburg, Germany}
\affiliation{University College London, London, United Kingdom}
\author{J.~Bj{\"o}rklund Svensson}
\affiliation{Deutsches Elektronen-Synchrotron DESY, Notkestr.~85, 22607 Hamburg, Germany}
\author{J.~M.~Garland}
\affiliation{Deutsches Elektronen-Synchrotron DESY, Notkestr.~85, 22607 Hamburg, Germany}
\author{P.~Gonz{\'a}lez Caminal}
\affiliation{Deutsches Elektronen-Synchrotron DESY, Notkestr.~85, 22607 Hamburg, Germany}
\affiliation{Universit{\"a}t Hamburg, Luruper Chaussee 149, 22761 Hamburg, Germany}
\author{B.~Hidding}
\affiliation{SUPA, Department of Physics, University of Strathclyde, Glasgow, United Kingdom}
\affiliation{The Cockcroft Institute, Daresbury, United Kingdom}
\author{G.~Loisch}
\affiliation{Deutsches Elektronen-Synchrotron DESY, Notkestr.~85, 22607 Hamburg, Germany}
\author{F.~Pe{\~n}a}
\affiliation{Deutsches Elektronen-Synchrotron DESY, Notkestr.~85, 22607 Hamburg, Germany}
\affiliation{Universit{\"a}t Hamburg, Luruper Chaussee 149, 22761 Hamburg, Germany}
\author{K.~P{\~o}der}
\author{S.~Schr{\"o}der}
\author{S.~Wesch}
\author{J.~C.~Wood}
\author{J.~Osterhoff}
\author{R.~D'Arcy}
\affiliation{Deutsches Elektronen-Synchrotron DESY, Notkestr.~85, 22607 Hamburg, Germany}

\date{\today}

\begin{abstract}
Energy-transfer efficiency is an important quantity in plasma-wakefield acceleration, especially for applications that demand high average power. 
Conventionally, the efficiency is measured using an electron spectrometer; an invasive method that provides an energy-transfer efficiency averaged over the full length of the plasma accelerator. 
Here, we experimentally demonstrate a novel diagnostic utilizing the excess light emitted by the plasma after a beam--plasma interaction, which yields noninvasive, longitudinally resolved measurements of the local energy-transfer efficiency from the wake to the accelerated bunch; here, as high as ($58\pm3$)\%. 
This method is suitable for online optimization of individual stages in a future multistage plasma accelerator, and enables experimental studies of the relation between efficiency and transverse instability in the acceleration process.
\end{abstract}

\maketitle
    
With the promise of gigavolts-per-meter accelerating gradients~\cite{LeemansNatPhys2006,HoganPRL2005,BlumenfeldNature2007}, plasma wakefields driven by intense laser~\cite{TajimaPRL1979} or particle beams~\cite{ChenPRL1985,RuthPA1985} offer a path toward compact accelerators for high-energy physics and photon science. 
In a beam-driven plasma-wakefield accelerator (PWFA), energy is transferred from a driver to the plasma by expelling plasma electrons from its path; some of this energy can subsequently be extracted by a trailing particle bunch placed at the correct distance behind the driver.
Trailing bunches with a sufficiently high current can alter the trajectory of the expelled plasma electrons as they return to the axis, modifying the local accelerating field while efficiently extracting energy from the wake~\cite{vanderMeerCLIC1985,ChenPRL1986,LotovPoP2005}.
This effect, known as \textit{beam loading} \cite{KatsouleasPA1986,TzoufrasPRL2008}, has been used to demonstrate energy-transfer efficiencies from the wake to the trailing bunch of up to ($42\pm4$)\% in PWFA, while simultaneously preserving beam properties such as charge and energy spread~\cite{LitosNature2014,LitosPPCF2016,LindstromPRL2021}.

Traditionally, energy-transfer efficiency is measured using magnetic dipole spectrometers by comparing the energy loss and gain of the driver and trailing bunch, respectively.
These measurements are longitudinally averaged over the full plasma stage, are invasive due to scattering of the charge by the diagnostic screen, and require stable, multi-shot measurements with varying beam imaging in order to ensure high energy resolution. 
As a result, experimental optimization is slow and precludes simultaneous spectrometer-based measurements of efficiency in individual stages of a future multistage plasma accelerator~\cite{SteinkeNature2016,LindstromPRAB2021}.

In this Letter, we present a novel method of measuring energy-transfer efficiency by detecting plasma-emission light from the dissipation of residual energy in the plasma wake.
Previous work has introduced plasma-emission light as a simple, nondestructive observable of the strength of a beam--plasma interaction, but was so far only used for alignment and synchronization~\cite{AdliNJP2016,Scherkl2022}.
Here, we not only observe an increased light yield in the presence of a beam driver, but also a reduction when a trailing bunch extracts energy from the wake.
By measuring the wake-energy-versus-plasma-light response curve, energy-transfer efficiency can thereby be inferred directly from the plasma light.
Importantly, this signal can be measured with high spatial resolution, enabling longitudinally resolved measurements of efficiency along the plasma accelerator.

\begin{figure*}[t]
  \includegraphics[width=\textwidth]{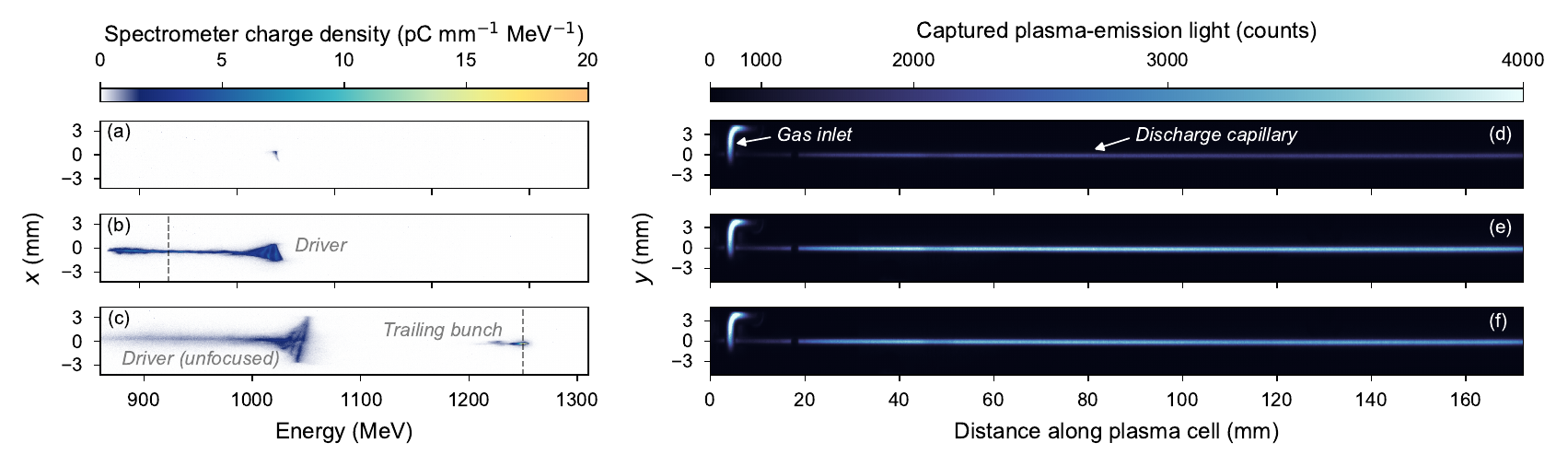}
  \caption{Measurements performed with a dipole spectrometer, showing three representative shots: (a) negligible beam charge; (b) beam driver only; and (c) both the driver and a trailing bunch. The imaging energy (dashed line) was optimized to resolve the relevant parts of the spectrum. Plasma-emission light signals corresponding to these spectra are shown in (d--f), where a quadratic color scale is used to highlight differences at high counts.}
  \label{fig:ExptSetup}
\end{figure*}

Observations of this effect took place at the FLASHForward plasma-accelerator facility at DESY \cite{DArcyRSTA2019}.
The FLASH linac \cite{SchreiberHPLSE2015} provided electron bunches from a photocathode, accelerated to a mean electron energy of 1052~MeV, compressed to a root-mean-square (rms) bunch length of 300~{\textmu}m, and linearly chirped in longitudinal phase space.
A triplet of energy collimators in a dispersive section provided control over the current profile \cite{SchroederIOP2020}; a notch collimator for generating a driver and trailing bunch, as well as head and tail collimators for adjusting their respective charges. 
The bunches were focused \cite{LindstromPRAB2020} at the entrance of a 195~mm-long, 1.5~mm-diameter capillary \cite{SpencePRE2000}, comprised of two milled sapphire slabs with gas inlets located 2.5~mm from either end.
Plasma was generated by discharging high-voltage pulses through the capillary, which was filled with argon gas (doped with 3\% hydrogen) at a pressure of 9~mbar.
By allowing the initial plasma to decay, the plasma density experienced by the bunches could be precisely controlled. 
The density in the flat-top region of the plasma 9.64~{\textmu}s after the start of the discharge was measured to be approximately $7\times10^{15}$~cm$^{-3}$ using an optical spectrometer to observe broadening of the H-alpha spectral line \cite{GigososSAB2003,GarlandRSI2021}.
Five quadrupole magnets were used to transport the outgoing electron bunches through a dipole spectrometer to a LANEX screen, onto which the beams were point-to-point imaged from the end of the plasma source [Figs.~\ref{fig:ExptSetup}(a--c)].
Finally, a CMOS camera provided spatially resolved measurements of the plasma-emission light radiated through one side of the plasma cell [Figs.~\ref{fig:ExptSetup}(d--f)].

An operating point with high energy-transfer efficiency [Fig.~\ref{fig:ExptSetup}(c)] was found by following the optimization technique described in Ref.~\cite{LindstromPRL2021}.
The mean energy of the driver electrons decreased by $\Delta E_{\mathrm{dec}} = (69 \pm 3)$~MeV, whereas the mean energy of the trailing-bunch electrons increased by $\Delta E_{\mathrm{acc}} = (194 \pm 4)$~MeV, as measured on the electron spectrometer, where the quoted uncertainty refers to the rms over the given shots.
The corresponding driver and trailing-bunch charges were $Q_{\mathrm{dec}} = (368 \pm 8)$~pC and $Q_{\mathrm{acc}} = (46.7 \pm 1.4)$~pC, respectively, based on the integrated charge density on the calibrated screen \cite{BuckRSI2010}.
The change in total energy of the bunches was therefore $\Delta W_{\mathrm{dec}} = Q_{\mathrm{dec}} \Delta E_{\mathrm{dec}} = (25.3 \pm 1.4)$~mJ and $\Delta W_{\mathrm{acc}} = Q_{\mathrm{acc}} \Delta E_{\mathrm{acc}} = (9.1 \pm 0.3)$~mJ, respectively.
Since the energy resolution of the electron spectrometer decreases away from the imaging energy, accurate spectral measurements of the two bunches cannot occur simultaneously. 
Therefore, separate measurements of the driver and the trailing bunch were performed, each with an optimized beam-imaging energy [Figs.~\ref{fig:ExptSetup}(b) and (c)].
The resulting energy-transfer efficiency, defined by
\begin{equation}
    \label{eq:espec_eff}
    \eta_s = \frac{\Delta W_{\mathrm{acc}}}{\Delta W_{\mathrm{dec}}},
\end{equation}
was calculated to be $(36 \pm 2)$\%, where the driver energy loss was assumed to remain constant during the trailing-bunch measurement.
In order to correct for imperfect charge coupling to the spectrometer, only shots with an accelerated trailing-bunch charge above 90\% of the measured maximum were considered.

Plasma-emission light was measured simultaneously to the spectral measurements, detected by the camera imaging the side of the plasma cell.
The integrated captured light was observed to vary with the incoming current profile, which was adjusted using the tail collimator in the dispersive section.
Beginning with the collimator completely inserted [Fig.~\ref{fig:ExptSetup}(a)], the light captured was primarily that from the discharge plasma with negligible beam interaction [Fig.~\ref{fig:ExptSetup}(d)].
Partially extracting the collimator until only the driver reached the plasma [Fig.~\ref{fig:ExptSetup}(b)] resulted in a substantial increase in the captured light [Fig.~\ref{fig:ExptSetup}(e)]---as the driver decelerated, energy was transferred to the plasma wake. 
This additional energy was eventually dissipated into the bulk plasma \cite{Zgadzaj2020,D'Arcy2022}, some of which was emitted as excess plasma-emission light.
When the collimator was further extracted such that a trailing bunch was introduced [Fig.~\ref{fig:ExptSetup}(c)], the excess plasma-emission light was observed to decrease  [Fig.~\ref{fig:ExptSetup}(f)].
This occurred because the trailing bunch extracted energy from the wake, making the reduction in plasma-emission light a simple and novel signature of the energy that remained in the plasma wake.

\begin{figure}[t]
  \includegraphics[width=\linewidth]{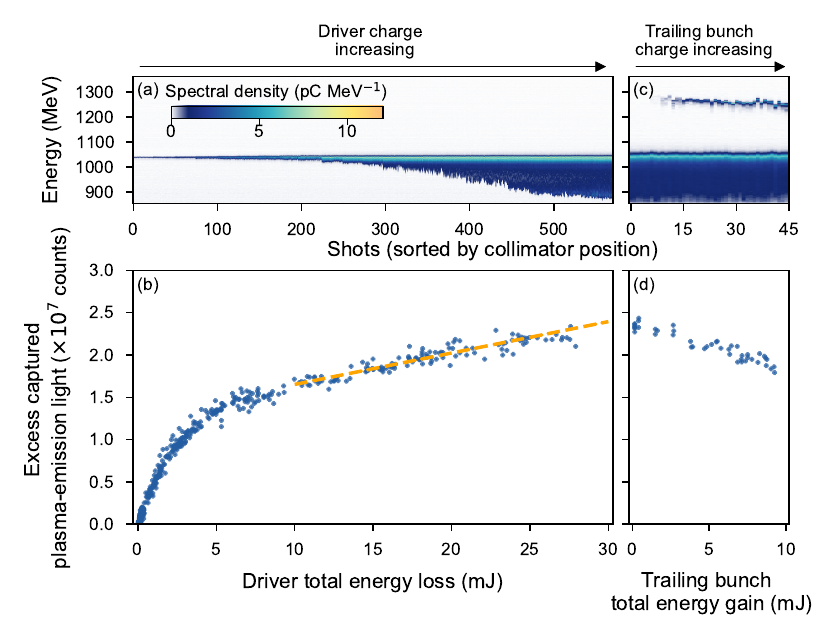}
  \caption{(a) Driver spectra measured as the collimator is extracted, introducing more charge to the plasma wake. (b) Plasma-emission light increases with increasing total driver energy loss (blue points), where the correlation is linear above 10~mJ (orange dashed trendline). (c) Further extracting the collimator, spectra of the trailing bunch show an increase in charge with the unfocused driver spectra remaining approximately constant. (d) The plasma-emission light decreases as the trailing bunch extracts energy from the plasma.}
  \label{fig:PlasmaLight}
\end{figure}

Figure~\ref{fig:PlasmaLight} demonstrates this effect in more detail.
By progressively increasing the amount of driver charge [Fig.~\ref{fig:PlasmaLight}(a)], the correlation between the total energy lost by the driver and the resulting excess light yield was established [Fig.~\ref{fig:PlasmaLight}(b)].
Here, the light is integrated spatially along the plasma accelerator, where each longitudinal segment is normalized to the light signal from a uniform plasma (as measured immediately after the discharge) in order to account for variations in light-collection efficiency across the field of view.
The response curve is affected by the specific conditions of the experiment: the light yield depends on the optical setup and the sensitivity of the camera chip, their spectral bandwidth and attenuation, the exposure-time settings, as well as on the initial state of the plasma before beam interaction.
Ultimately, this calibration allows us to estimate how much excess energy is deposited in the plasma wake based solely on the plasma-emission light.

Gradually introducing the trailing bunch [Fig.~\ref{fig:PlasmaLight}(c)], plasma-emission light was measured as a function of the total energy gained [Fig.~\ref{fig:PlasmaLight}(d)]. 
In order to accurately quantify this extracted energy, all plasma-interacted charge must be transported to the spectrometer.
To reduce the systematic error due to imperfect charge coupling, only shots with more than 90\% of the maximum measured trailing-bunch charge at each collimator position are considered.
The trailing bunch extracted a maximum of 9.3~mJ from the wakefield.
In the corresponding region of the light-versus-energy-loss measurement [above $1.65\times10^7$ counts in Fig.~\ref{fig:PlasmaLight}(b)], a linear fit can be used to model the plasma-emission light yield $I_p = f(\Delta W_{\mathrm{wake}})$, where $\Delta W_{\mathrm{wake}}$ is the energy deposited in the plasma wake, equal to $\Delta W_{\mathrm{dec}}$ when only the driver is present.
The correlation in Fig.~\ref{fig:PlasmaLight}(d) has an approximately equal and opposite slope to that of the model $f(\Delta W_{\mathrm{wake}})$, suggesting that the excess plasma-emission light is independent of how the wake energy reaches its final value---whether the wake has purely seen energy deposition from a driver, or also extraction by a trailing bunch. 
As a consequence, the remaining wake energy can be estimated from the excess plasma-emission light by inverting the previously established model
\begin{equation}
    \label{eq:plasma-light-wake-energy}
    \Delta W_{\mathrm{wake}} = \Delta W_{\mathrm{dec}} - \Delta W_{\mathrm{acc}} = f^{-1}(I_p).
\end{equation}    

Equation~(\ref{eq:plasma-light-wake-energy}) provides an alternative way of calculating the energy-transfer efficiency.
The total energy gained by the trailing bunch can be inferred from the plasma-emission light via $\Delta W_{\mathrm{acc}} = \Delta W_{\mathrm{dec}} - f^{-1}(I_p)$, which requires a measurement of the response curve $f$ [Fig.~\ref{fig:PlasmaLight}(b)] and the total driver energy loss. 
Substituting this into Eq.~(\ref{eq:espec_eff}), the energy-transfer efficiency can be expressed as
\begin{equation}
    \label{eq:plasma_eff}
    \eta_p = 1 - \frac{f^{-1}(I_p)}{\Delta W_{\mathrm{dec}}}.
\end{equation}

For an accurate comparison of the plasma-light-based measurement [Eq.~(\ref{eq:plasma_eff})] and the spectrometer-based measurement [Eq.~(\ref{eq:espec_eff})], the energy spectrum of the trailing bunch must be accurately measured.
This requires point-to-point imaging of the trailing bunch, meaning that the spectrum of the driver cannot be simultaneously measured with high accuracy.
Nevertheless, the total driver energy loss can be estimated from the mean plasma-emission light yield when only the driver is present $\langle I_{p} \rangle_0$ [leftmost shots in Fig.~\ref{fig:PlasmaLight}(c--d)], where $\langle \rangle$ denotes an average over multiple shots.
Further improving this estimate, drifts and correlations can be accounted for by considering the unfocused spectrum of the driver.
While the \textit{absolute} driver spectrum is not perfectly accurate in this measurement, due to reduced charge transmission and energy resolution,
we only need to consider \textit{relative} changes to the driver spectrum. If we assume that the spectrum approximately retains its normalized shape (i.e., charge is lost approximately uniformly across the spectrum), the relative changes in driver charge and mean energy loss (compared to the averages measured for the driver-only shots) can be used to correct for shot-to-shot fluctuations.
The resulting best estimate of the total energy loss of the driver for each shot is therefore
\begin{equation}
    \label{eq:driver-energy-loss-estimate}
    \Delta W_{\mathrm{dec}} \approx f^{-1}(\langle I_{p} \rangle_0) \frac{Q_{\mathrm{dec}} \Delta E_{\mathrm{dec}}}{\langle Q_{\mathrm{dec}} \Delta E_{\mathrm{dec}}\rangle_0}.
\end{equation}
For consistency, the same value of $\Delta W_{\mathrm{dec}}$ will be used in Eqs.~(\ref{eq:espec_eff}) and (\ref{eq:plasma_eff}) when comparing the two methods.

\begin{figure}[t]
  \includegraphics[width=\linewidth]{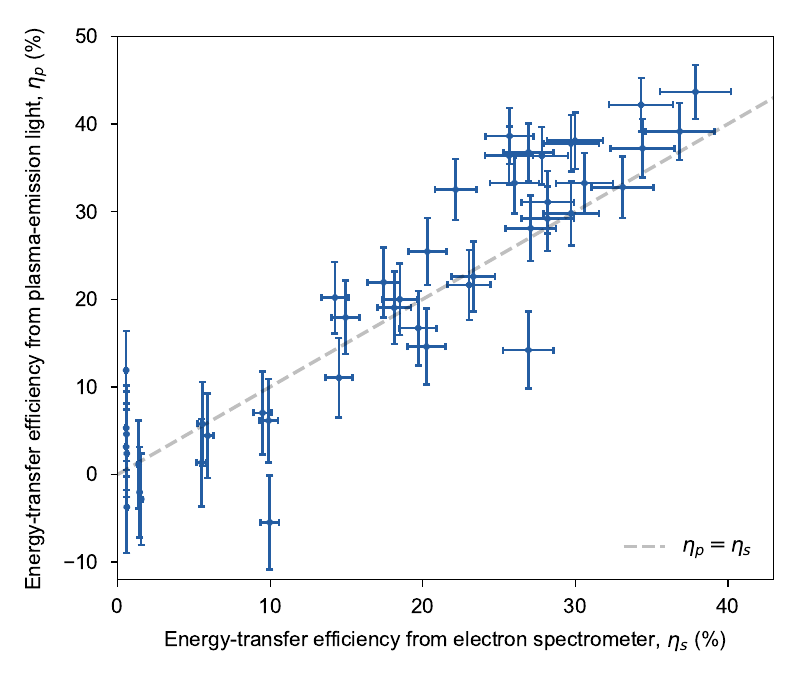}
  \caption{Comparison of the longitudinally averaged energy-transfer efficiency measured using the plasma-emission light and the electron spectrometer, considering the shots shown in Figs.~\ref{fig:PlasmaLight}(c--d).
  The majority of the data points show agreement between the two methods, i.e., $\eta_p = \eta_s$ (gray dashed line), within the error of each measurement.
  Negative efficiencies here are due to shot-to-shot jitter in driver energy loss, resulting in some shots with $I_p$ lower than $\langle I_p \rangle_0$.}
  \label{fig:EfficiencyComp}
\end{figure}

Figure~\ref{fig:EfficiencyComp} shows such a comparison of energy-transfer efficiency.
Using the shots in Fig.~\ref{fig:PlasmaLight}(c--d), the two methods agree to within the measurement error (approximately 4 percentage points).
The main sources of error in both measurements arise from the fit error of the model $f$, as well as the shot-to-shot jitter of the total driver energy loss [Eq.~(\ref{eq:driver-energy-loss-estimate})].
The standard deviation of both $\langle Q_{\mathrm{dec}} \Delta E_{\mathrm{dec}}\rangle_0$ and $\langle I_{p} \rangle_0$ were used to quantify their contributions to the overall uncertainty.

Having established the agreement between the spectrometer-based and plasma-light-based efficiency measurements, the capabilities of each approach can be compared. 
While electron spectrometers remain vital components of any plasma accelerator and are indeed required to initially construct the wake-energy-versus-plasma-light response curve [Fig.~\ref{fig:PlasmaLight}(b)], the measurement is inherently invasive because the beams must be strongly dispersed as well as passed through a screen.
The plasma-light-based method, on the other hand, is noninvasive.
More importantly, whereas the spectrometer measures the longitudinally averaged efficiency and is affected by charge loss at any point during the acceleration process, the plasma light is measured along the full length of the plasma accelerator. This implies that the local energy-transfer efficiency can be calculated with longitudinal resolution, unaffected by downstream loss of charge.

\begin{figure*}[t]
  \includegraphics[width=0.75\textwidth]{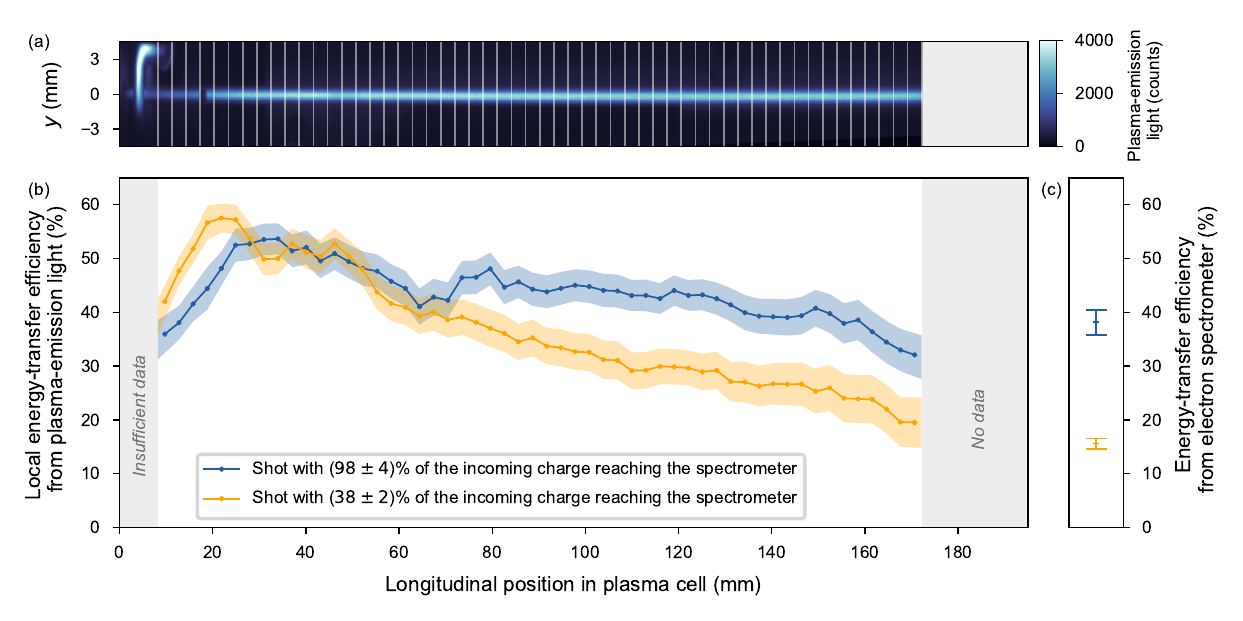}
  \caption{(a) The plasma emission-light is separated into 3 mm-long segments (gray lines). 
  (b) Local energy-transfer efficiency is shown as a function of position along the plasma cell.
  Two separate shots were considered: one with full charge coupling (blue line), and another with lower charge coupling (orange line). 
  The error bands represent the uncertainty of the response-curve fit for each segment, as well as the jitter of the total driver energy loss.
  (c) Corresponding spectrometer-based measurements of efficiency are shown.}
  \label{fig:SpatialResolution}
\end{figure*}

In order to longitudinally resolve the energy-transfer efficiency, the plasma-light signal can be divided into segments [Fig.~\ref{fig:SpatialResolution}(a)]. 
Each segment will have its own response curve, similar to that measured for the full plasma stage in Fig.~\ref{fig:PlasmaLight}(b), which depends on the driver energy loss within that segment.
This relation could not be directly determined in this experiment, as only the integrated energy loss was measured by the spectrometer.
However, we expect the local energy loss to be approximately proportional to the integrated energy loss.
Under this assumption, the measurement of the local excess light as a function of the integrated energy loss can be used as the response curve for a given segment.
This is sufficient for calculating the energy-transfer efficiency, as Eq.~(\ref{eq:plasma_eff}) is invariant to scaling of the response curve $f$ when substituting in Eq.~(\ref{eq:plasma-light-wake-energy}).
Note that an error in the above assumption will simply imprint a systematic trend that affects all shots equally, meaning that a comparison of the local energy-transfer efficiency between shots is valid regardless---a key feature for online optimisation.

Figure~\ref{fig:SpatialResolution}(b) demonstrates a longitudinally resolved measurement of energy-transfer efficiency for two separate shots with an accelerated trailing bunch. 
Both measurements show a rapid initial increase in energy-transfer efficiency, likely due to a plasma-density upramp at the entrance of the cell, followed by a gradual decrease throughout the uniform-density region of the plasma, possibly caused by an evolution of the wakefield generated by the driver.
However, although both shots had approximately the same input conditions, including an incoming trailing-bunch charge of ($50\pm2$)~pC, clear differences can be observed.
A shot with full charge coupling (49~pC on the spectrometer) displays a local energy-transfer efficiency initially reaching 54\% followed by a slow decrease, at a rate of 0.11\%~mm$^{-1}$, resulting in an efficiency measured at the spectrometer of 38\% [Fig.~\ref{fig:SpatialResolution}(c)].
Another shot, which had a significantly lower charge coupling (19~pC or 38\% of the incoming charge), while at first reaching a higher efficiency of 58\%, displays a more rapid decline throughout the cell, at a rate of 0.23\%~mm$^{-1}$, and an efficiency of only 16\% measured at the spectrometer.
This higher rate of decline within the plasma is likely caused by a continuous loss of charge starting at 50--70~mm into the plasma cell, while the further reduction at the spectrometer can be explained by an additional loss in the post-plasma transport.
Ultimately, the evolution of the energy-transfer efficiency in this second shot, which appears identical in terms of measured input parameters, is consistent with the growth of a transverse instability.
We speculate that this is connected to the high initial efficiency \cite{LebedevPRAB2017}---an important effect that, using this method, can now be studied experimentally.

Looking ahead, the noninvasive nature of this method will enable online optimization and monitoring of plasma accelerators also when the trailing bunch cannot be intercepted by a spectrometer.
This applies during operation in most applications, and in particular for a multistage plasma accelerator \cite{PeiPAC2009,AdliSLAC2013}, where separate diagnosis of the acceleration process in each stage will be crucial, but cannot be provided by a single spectrum measurement.

In summary, a novel method for measuring the energy-transfer efficiency in a plasma accelerator was presented, using excess plasma-emission light to infer the energy remaining in the wake. 
This enabled the longitudinally resolved measurement of the local energy-transfer efficiency evolution throughout the acceleration process.
We foresee that this simple, noninvasive technique can provide deeper understanding and more reliable operation of plasma-accelerator stages---especially important as the field looks toward multiple stages and applications.

\begin{acknowledgments}
    The authors would like to thank M.~Dinter, S.~Karstensen, S.~Kottler, K.~Ludwig, F.~Marutzky, A.~Rahali, V.~Rybnikov, A.~Schleiermacher, the FLASH management, and the DESY FH, FS, M and V divisions for their scientific, engineering and technical support. This work was supported by Helmholtz ARD, as well as the Maxwell computational resources at DESY.
\end{acknowledgments}


\end{document}